\documentclass{article} 
\usepackage{iclr2022_conference,times}

\usepackage{amsmath,amsfonts,bm}

\def\eqref#1{equation~\ref{#1}}

\def\1{\bm{1}}

\def\rd{{\textnormal{d}}}

\def\vu{{\bm{u}}}
\def\vv{{\bm{v}}}

\DeclareMathAlphabet{\mathsfit}{\encodingdefault}{\sfdefault}{m}{sl}
\SetMathAlphabet{\mathsfit}{bold}{\encodingdefault}{\sfdefault}{bx}{n}

\def\gD{{\mathcal{D}}}

\def\gK{{\mathcal{K}}}

\def\gN{{\mathcal{N}}}

\def\gQ{{\mathcal{Q}}}
\def\gR{{\mathcal{R}}}

\newcommand{\E}{\mathbb{E}}

\newcommand{\R}{\mathbb{R}}

\newcommand{\softmax}{\mathrm{softmax}}

\usepackage{mathtools}
\usepackage{multirow}
\usepackage{multicol}
\usepackage{booktabs}
\usepackage{subfigure}

\DeclareMathOperator*{\uniform}{Uniform}
\DeclareMathOperator*{\bmsample}{BM25}

\DeclareMathOperator*{\transformerenc}{Trans-Enc}

\usepackage{hyperref}
\usepackage{url}
\usepackage{color}

\title{Towards Robust Ranker for Text Retrieval}

\iclrfinalcopy

\author{
Yucheng Zhou$^1$\thanks{Equal contribution.}~~\thanks{Work is done during internship at Microsoft.}~, Tao Shen$^{2*}$, Xiubo Geng$^2$, Chongyang Tao$^2$, Can Xu$^2$, \\
\textbf{~Guodong Long$^1$, Binxing Jiao$^2$, Daxin Jiang$^2$}\thanks{Corresponding author.} \\
$^1$AAII, School of CS, FEIT, University of Technology Sydney \\
\texttt{yucheng.zhou-1@student.uts.edu.au, guodong.long@uts.edu.au} \\
$^2$Microsoft \\
\texttt{\{shentao,xiubo.geng,chongyang.tao,can.xu\}@microsoft.com} \\
\texttt{\{binxjia,djiang\}@microsoft.com}
}

\begin{document}

\maketitle

\begin{abstract}
A ranker plays an indispensable role in the de facto `retrieval \& rerank' pipeline, but its training still lags behind -- learning from moderate negatives or/and serving as an auxiliary module for a retriever. In this work, we first identify two major barriers to a robust ranker, i.e., inherent label noises caused by a well-trained retriever and non-ideal negatives sampled for a high-capable ranker. Thereby, we propose multiple retrievers as negative generators improve the ranker's robustness, where i) involving extensive out-of-distribution label noises renders the ranker against each noise distribution, and ii) diverse hard negatives from a joint distribution are relatively close to the ranker's negative distribution, leading to more challenging thus effective training. To evaluate our robust ranker (dubbed \textsc{R$^2$anker}), we conduct experiments in various settings on the popular passage retrieval benchmark, including BM25-reranking, full-ranking, retriever distillation, etc. The empirical results verify the new state-of-the-art effectiveness of our model. 
\end{abstract}

\section{Introduction} \label{sec:intro}
Text Retrieval plays a crucial role in many applications, such as web search \citep{Brickley19Google} and recommendation \citep{Zhang19Deep}. 
Given a query, it aims to retrieve all relevant documents from a large-scale collection (while each entry of the collection could be a sentence, passage, document, etc., we adopt `document' for a clear demonstration).
For a better efficiency-effectiveness trade-off, the \textit{de facto} paradigm relies on a `retrieval \& rerank' pipeline \citep{Guo22Semantic}. 
That is, `retrieval' is to use an efficient retriever to fetch a set of document candidates given a query, whereas `rerank' is to re-calculate the relevance of the query to each candidate by a heavy yet effective ranker for better results. 

\begin{figure}[t]
    \centering
    \begin{minipage}[t]{\linewidth}
    \centering
    \includegraphics[width=0.12\linewidth]{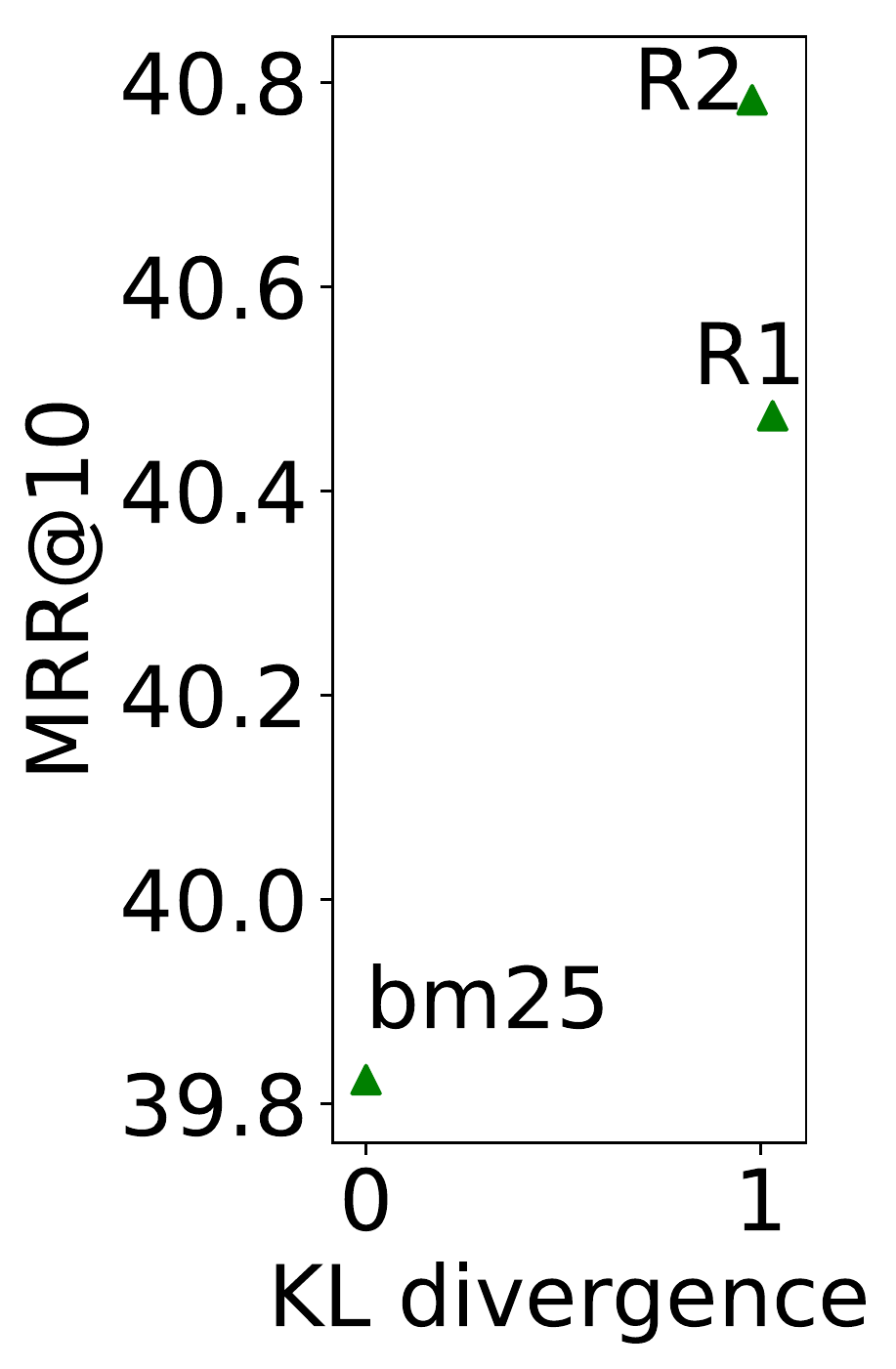}
    \includegraphics[width=0.86\linewidth]{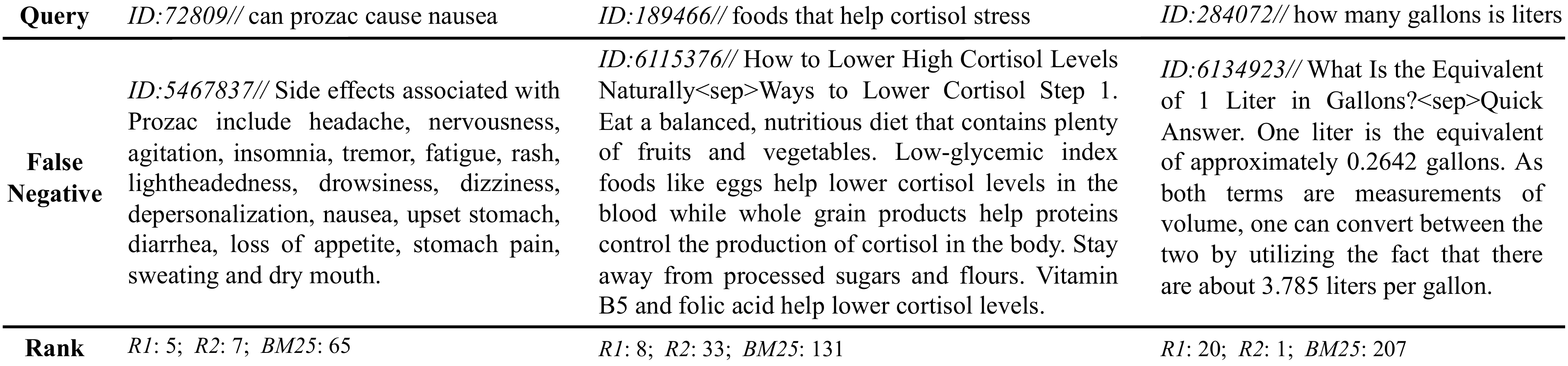}
    \caption{\small 
    BM25-reranking performance of the rankers trained on different negative distributions by specific retrievers (left) and false negative labels brought by the two well-trained strong retrievers in contrast to BM25 retriever (right). Here, `R1' denotes a well-trained coCondenrser \citep{Gao22Unsupervised} dense-vector retriever whereas `R2' denotes a well-trained SPLADE \citep{Formal21SPLADEv2} lexicon-weighting retriever. 
    }
    \label{fig:intro}
    \end{minipage}
\end{figure}

Although many recent works are presented to improve the retrievers, how to learn an effective ranker remains a long-term open question. 
Distinct from categorical classification \citep{Zhang15Character} or semantic relatedness \cite{Gabrilovich07Computing} tasks, only positive query-document pairs are provided in retrieval tasks, so a critical prerequisite of training a ranker is sampling negative documents from the collection for training queries. 
A brute-force but widely-adopted way is to learn a ranker with off-the-shelf BM25 negatives, but the moderate negatives cannot challenge a high-capable ranker built upon pre-trained language models (PLM) with cross-encoder structure, leading to sub-optimal results (see the BM25 point in Figure~\ref{fig:intro}(left)). 

A recent trend is resorting to a retriever to sample hard negatives from the collection for ranker training in an adversarial \citep{Zhang21Adversarial} or cooperative \citep{Ren21RocketQAv2} manner  (see the other two points in Figure~\ref{fig:intro}(left)). 
That is, compared with BM25 negatives, a PLM-based lightweight retriever can sample relatively hard negatives to challenge the ranker, and in turn, the stronger ranker can provide more effective distillation signals for retriever training. 
However, a strong well-trained retriever as the negative sampler is more likely to introduce label noises, a.k.a. false negative labels in retrieval field \citep{Qu21RocketQA}. 
Please refer to Figure~\ref{fig:intro}(right) for several examples. 
This problem is caused by, when applying finite manpower to infinite documents in the collection, the annotating process depends heavily on the best on-hand retriever to narrow the labeling range.

On the other hand, since a ranker is usually built upon a PLM-based cross-encoder without consideration of computation complexity, it is far more high-capable than any retriever with bi-encoder. As such, the hard negatives sampled by a single retriever hardly fool the ranker, making the ranker training less effective. 

To reduce the effect of the two challenges above, we propose a brand-new robust reranker (\textsc{R$^2$anker}) learning method. 
It involves multiple retrievers as generators to jointly sample diverse hard negatives, which are used to challenge a single ranker as the discriminator. 

\textsc{R$^2$anker} has certain merits regarding the robustness of its model training.
First, as the false negatives are closely subject to the relevance distribution over the collection by a specific retriever, various negative generators achieved by different retrievers are prone to sample out-of-distribution (OOD) or open-set label noise \cite{Wei21Open} to each other. In light of `\textit{insufficient capacity}' assumption \citep{Arpit17Closer}, such open-set noise has been proven effective in improve robustness \citep{Wei21Open} when learning a ranker with open-set noises. Second, intuitively, sampling negative over a joint distribution of various retrievers is likely to offer more challenging hard negatives. Although the joint distribution is not exactly same as the ranker's negative distribution (which is unavailable due to combinatorial explosion), at least it is closer to the ideal negative distribution than a single retriever verified in our analysis. 
Hence, the ranker can learn from diverse hard negatives that are approximatively subject to ideal negative distribution for more effectiveness. 

In experiments, we adopt the most popular passage retrieval benchmark dataset, MS-Marco \citep{Nguyen16Tarek}, to evaluate our proposed model under various settings. 
Specifically, our method achieves new state-of-the-art performance on BM25 reranking and full-ranking. Meantime, to verify the expressive power of our ranker model, we conduct an experiment to distill our well-trained model to a retriever, which even shows state-of-the-art first-stage retrieval performance without cooperative training. 
Moreover, our extensive quantitative analyses also unveil the essence regarding negative distributions to reach a robust ranker. 
We make our code implementations and well-trained models available at \url{https://github.com/taoshen58/R2ANKER}.

\section{R$^2$anker: Robust Ranker} \label{sec:method}

\subsection{Task Formulations \& Challenges}  \label{sec:formulation_challenges}

Given a user query $q$, a ranker model, $\gK(q,d)$, is responsible for calculating a relevance score between $q$ and an arbitrary document $d$ from a large-scale collection $\gD$ (i.e., $d\in\gD$). 
It usually serves as a downstream module for an efficient retriever, $\gR$, to compose a `retrieval \& rerank' pipeline, where a lightweight retriever $\gR$ (e.g., Siamese encoder) is to retrieve top candidates and then a relatively heavy $\gK$ (says cross-encoder \citep{Devlin19BERT}) to make the search quality better. 

Formally, $\gK(q,d)$ is usually achieved by a deep Transformer encoder for dense interactions in a pair of query and document (so called cross-encoder, or one-stream encoder), i.e., 
\begin{align}
    \gK(q,d) = \transformerenc(\texttt{[CLS]} q \texttt{[SEP]} d \texttt{[SEP]}; \theta^{(ce)}) \in\R.
\end{align}
In contrast, a retriever $\gR(q,d)$ is usually defined as a bi-encoder (a.k.a. dual-encoder, two-stream encoder, and Siamese encoder) to derive counterpart-agnostic representation vectors, i.e., 
\begin{align}
    &\gR(q,d) = \vu^T\vv, \\
    \notag ~~\text{where,}~\vu = \transformerenc(\texttt{[CLS]} q &\texttt{[SEP]}; \theta^{(be)}),  \vv = \transformerenc(\texttt{[CLS]} d \texttt{[SEP]}; \theta^{(be)}) \in\R^d.
\end{align}

Distinct from a traditional classification task where each category is associated with adequate training samples, only positive document(s), $d^+$, is provided for each training query $q\in\gQ^{\text{(trn)}}$, regardless of its negative ones, $d^-$. Therefore, to train a ranker, a prerequisite is determining a negative sampling strategy to make the training procedure more effective, i.e., 
\begin{align}
    \gN = \{d | d \sim P(\gD \backslash \{d^+\}|q; \theta^{\text{(smp)}}) \}, \label{eq:basic_hn_sampling}
\end{align}
where $P$ denote a probability distribution over $\gD$, which can be either non-parametric (i.e., $\theta^{\text{(smp)}}=\emptyset$) or parametric (i.e., $\theta^{\text{(smp)}}\neq\emptyset$). 
Then, the ranker calculates a probability distribution over a combination of $\{d^+\}$ and $\gN$, i.e.,
\begin{align}
    P(\rd|q; \theta^{\text{(ce)}}) =  \dfrac{\exp(\gK(q,d))}{\sum_{d\in\{d^+\}\cap\gN} \exp(\gK(q,d))},~\text{where,}~d\in \{d^+\}\cap\gN. \label{eq:basic_contrastive}
\end{align}
Lastly, the ranker is trained via a contrastive learning objective, whose training loss is defined as
\begin{align}
    L_{\theta} \!=\! - \!\sum\nolimits_{q, d^+}\!\! \log P(\rd=d^+|q; \theta^{\text{(ce)}}),~\text{where,}~\rd\in \{d^+\}\cap\gN. \label{equ:gene_contrast}
\end{align}

Nonetheless, there are \textit{two major challenges} emerging in ranker training: 
First, as crowd-sourcing is very expensive, it is possible to annotate every positive document $d^+$ for a query $q$, leading to a label-noise problem. That is, there exists false negative documents in $\gD\backslash\{d^+\}$. What's worse, the false negatives are more likely to fuse with hard negatives (i.e., with high values of $P(\gD\backslash\{d^+\}|q$), making the ranker training non-robust.

Second, as exhaustive training (i.e., $\gN=\gD\backslash\{d^+\}$) is infeasible in practice, how to train the model effectively with limited computation resources. 
Ideally, the model to sample negatives for a query should rely on the being trained model to make the learning more effective, i.e., $\theta^{\text{(smp)}}$ in Eq.(\ref{eq:basic_hn_sampling}) equaling to $\theta^{\text{(ce)}}$ of $\gK$.
However, $\gK$ as a sampler over $P(\gD\backslash\{d^+\}|q; \theta^{\text{(ce)}})$ suffers from a combinatorial explosion problem brought from the cross-encoder, leading to intractable computation overheads. 
Practically, $\theta^{\text{(smp)}}$ is must as efficient as possible to circumvent the problem, which could be a heuristic strategy (e.g., uniform sampling), lightweight term-based retriever (e.g., BM25), or later-interaction representation models (e.g., Siamese encoder). 
So, it requires a sophisticated negative sampling strategy to simulate $P(\gD\backslash\{d^+\}|q; \theta^{\text{(ce)}})$ during ranker training.

\subsection{Brute-Force: Random \& BM25 Negatives}

The most straightforward negative sampling method to fulfill Eq.(\ref{eq:basic_hn_sampling}) is uniform sampling over the whole collection $\gD$ (a.k.a random or in-batch negatives). That is 
\begin{align}
    \gN^{\text{(rdm)}} = \{d | d \sim \uniform(\gD \backslash \{d^+\}|q)\}. \label{eq:hn_random}
\end{align}
Even if using large numbers of random negatives in one batch is proven very effective in self-supervised contrastive learning, prior works show very inferior retrieval quality when only random negatives were involved. 
The common practice is resorting to a lightweight lexicon-based retriever, i.e., BM25, to sample relatively challenging negatives (a.k.a. hard negatives) for a specific query. This is written as
\begin{align}
    \gN^{\text{(bm25)}} = \{d | d \sim \bmsample(\gD \backslash \{d^+\}|q; \gD) \}, \label{eq:hn_bm25}
\end{align}
where BM25 model is built upon the collection $\gD$. 

As \textit{BM25-reranking}\footnote{A BM25 retriever fetches top (says 100 or 1000) candidates from a large-scale collection, and then a ranker is asked to assign a relevance score for better search results.} usually serves as a critical task to evaluate a trained retriever, learning from BM25 negatives, $\gN^{\text{(bm25)}}$, is closely consistent between the training and evaluating phase, potentially leading to optimal evaluation results. 

However, both challenges mentioned in \S\ref{sec:formulation_challenges} remain unsolved. 
Limited by discrepancy against ranker distribution and label-noise training data, training a ranker based solely on BM25 negatives, $\gN^{\text{(bm25)}}$, is far from optimal performance and leaves a large room to improve. 

\subsection{Effective Ranker with Adversarial Hard Negatives}

To alleviate the first challenge, a well-known method is to sample hard negatives by applying a strong or well-train retriever, $\gR$, to a query $q$, which is parameterized by $\theta^{\text{(be)}}$. That is,
\begin{align}
    \gN^{\text{(be)}} = \{d |d \sim P(\gD\backslash\{d^+\}|q; \theta^{\text{(be)}})\}.
\end{align}
Although $P(\gD\backslash\{d^+\}|q; \theta^{\text{(be)}})$ is not very similar to the ideal distribution $P(\gD\backslash\{d^+\}|q; \theta^{\text{(ce)}})$, the strong deep retriever at-least ensure its distribution is closer to the ideal one than BM25, thus leading to better performance in empirical. 

Then, the hard negatives are used to minimize the contrastive learning loss defined in Eq.(\ref{eq:basic_contrastive}).
Coupled with a recently advanced adversarial retriever-ranker learning method \citep{Zhang21Adversarial}, the learning objective can be written as
\begin{align}
    J^{\gR^*, \gK^*} \!=\! \min\nolimits_{\theta^{\text{(be)}}} \max\nolimits_{\theta^{\text{(ce)}}} \E_{\gN^{\text{(be)}} = \{d |d \sim P(\gD\backslash\{d^+\}|q; \theta^{\text{(be)}})\}}
    \left[ \log P(\rd=d^+|q; \theta^{\text{(ce)}}) \right]. \label{eq:adv_train}
\end{align}
where the $\theta^{\text{(be)}}$-parameterized $\gR$ can be either a frozen and well-trained or a jointly optimized retriever.

\subsection{Robust Ranker with Open-Set Diverse Negative}

Although learning from (adversarial) hard negatives has been proven effective to obtain a high-performing ranker \citep{Ren21RocketQAv2,Zhang21Adversarial}, a single retriever $\gR$, even well-trained with various advanced techniques \citep{Qu21RocketQA,Lu22ERNIE}, is hard to provide hard enough negatives to challenge the ranker $\gR$ for robust training.

What's worse, a strong well-trained retriever usually introduces the label-noise problem. 
Due to limited crowd-sourcing resources, it is impossible to comprehensively annotate the relevance of every query $\forall q\in\gQ^{\text{(trn)}}$ to every document $\forall d\in\gD$. 
In general, the annotating process can be roughly described as 
i) using the best on-hand retriever (e.g., a commercial search engine) to fetch top document candidates for a query $q$, and then ii) distinguish positive document(s), $d^+$, associated to $q$ from the very top candidates. 
Therefore, constrained by the retriever in the annotation process, there exists positive documents for $q$ not included in the top candidates, which are regarded as negative by mistake -- false negative label -- degrading standard ranker training. Prior works focus on `co-teaching' or/and `boosting' strategies \citep{Qu21RocketQA,Zhang21Adversarial}, but they assume a ranker is robust enough for anti-noise while only denoise for more fragile retrievers by the ranker. 
Therefore, how to 

From another point of view, we could also formulate the search problem (both retrieval and rerank) as a many-class many-label classification problem, where the number of classes equals to $|\gD|$, i.e., the number of documents in $\gD$. And $|\gD|$ is usually very large, ranging from millions to billions. Thus, the current solutions of the search problem are analogous to \textit{label semantic matching} paradigm for many-class classification problems \citep{Hsu19Multi}. 
As such, the mis-labeled class caused by a single $\theta^{\text{(be)}}$-parameterized retriever $\gR$ will be subject to the following distribution:
\begin{align}
    y' \sim P^{\text{(FN)}}(\gD\backslash\{d^+\}|q; \theta^{\text{(be)}}), \label{eq:noise_fn_main}
\end{align}
where $P^{\text{(FN)}}(\cdot|\cdot;\theta^{\text{(be)}})$ denotes an inherent label noise distribution by the retriever $\theta^{\text{(be)}}$. 

Motivated by weak negatives and label noises, we propose to leverage multiple retrievers to improve the ranker's robustness from two aspects -- i) introducing out-of-distribution noise against inherent label noise and ii) generating diverse hard negatives to challenge the only ranker. 

As proven by a recent ``\textit{insufﬁcient capacity}'' \citep{Arpit17Closer} assumption\footnote{As stated in \citep{Wei21Open}, ``increasing the number of examples while keeping representation capacity fixed would increase the time needed to memorize the data set. Hence, the larger the size of auxiliary dataset is, the more time it needs to memorize the open-set noises in the auxiliary dataset as well as the inherent noises in the training set, relative to clean data.''}, learning extra out-of-distribution (OOD) or open-set noise can improve robustness against inherent label noises that are subject to one dataset or one distribution. 
Inspired by this, we present one (or several) extra retriever (parameterized by $\tilde\theta^{\text{(be)}}$) other than that in Eq.(\ref{eq:noise_fn_main}) to introduce OOD documents as open-set noise labels and alleviate the effect of noisy training samples $(q, y')$. 
In formal, the open-set label noise can be generated by
\begin{align}
    \tilde y' \sim P^{\text{(FN)}}(\gD\backslash\{d^+\}|q; \tilde\theta^{\text{(be)}}),
\end{align}
which will be mixed with $(q, y')$ to learn a more robust ranker. 

Meantime, employing multiple retrievers can provide more diverse hard negatives -- seen as different negatives distributions -- prone to be more challenging for ranker training -- leading to robust ranker. Intuitively, the joint negative distribution of multiple retrievers is more close to the ideal negative distribution subjecting to $\theta^{\text{(ce)}}$ as top-probable negatives cannot be handled by most retrievers. 

Consequently, we can reformulate Eq.(\ref{eq:adv_train}) as 
\begin{align}
    &J^{\gK^*} \!=\! \min\nolimits_{\Theta^{\text{(be)}}} \max\nolimits_{\theta^{\text{(ce)}}} \E_{ \gN^{\text{(be)}} = \{d |d \sim P(\gD\backslash\{d^+\}|q; \Theta^{\text{(be)}})\}}
    \left[ \log P(\rd=d^+|q; \theta^{\text{(ce)}}) \right], \label{eq:adv_train_final} \\
    &~~\text{where,}~P(\gD\backslash\{d^+\}|q; \Theta^{\text{(be)}}) = \prod\nolimits_{\theta^{\text{(be)}}_k \in \Theta^{\text{(be)}}} P(\gD\backslash\{d^+\}|q; \theta^{\text{(be)}}_k), \Theta^{\text{(be)}} = \{\theta^{\text{(be)}}_1, \theta^{\text{(be)}}_2, \dots \}. \label{eq:neg_distribution_final}
\end{align}
Here, each $\theta^{\text{(be)}}_k$ parameterizes a retriever (if applicable), which can be any high-efficient model. 

\subsection{Efficient Implementations}

To ground our proposed robust ranker, several details must be specified before our experiments.

First, as the success of our robust rank training depends on the multiple retrievers' diversity in Eq.(\ref{eq:neg_distribution_final}), we need to employ multiple retrievers distinct enough to each other. 
To this end, we employ three kinds of retrievers with five retrieval models in total:
\begin{itemize}
\item i) \textbf{BM25} retriever: A simple BM25 retrieval model built on the whole collection. 
\item ii) \textbf{Den-BN} retriever: A dense-vector retrieval model trained on \textit{BM25} negatives. 
\item iii) \textbf{Den-HN} retriever: A dense-vector retrieval model that is trained on hard negatives sampled by \textit{Den-BN}. 
\item iv) \textbf{Lex-BN} retriever: A lexicon-weighting retrieval model trained on \textit{BM25} negatives. 
\item v) \textbf{Lex-HN} retriever: A lexicon-weighting retrieval model that is trained on hard negatives sampled by \textit{Lex-BN}. 
\end{itemize}
In our experiments, we employ two state-of-the-art retriever for dense-vector retrieval \citep{Gao22Unsupervised} and lexicon-weighting retrieval \citep{Formal21SPLADEv2}, respectively. 

Second, in line with \citep{Clark20ELECTRA} and \citep{Ren21RocketQAv2}, we do not seek for updating the generators (i.e., the retrievers in our method) w.r.t performance of the discriminator (i.e., the ranker) with two considerations: On the one hand, we try to avoid heavy computation overheads to train the retrievers jointly and update the large-scale index synchronously. On the other hand, due to their intrinsic discrepancy in model structure, the generators hardly fool the discriminator, making the adversarial process less effective. As verified in \citep{Zhang21Adversarial}, cooperative learning (training the retriever towards the reranker, regularized by a Kullback–Leibler divergence) is also necessary for competitive performance. 

Third, the strategy to sample a negative distribution in Eq.(\ref{eq:basic_hn_sampling}) plays an important role in our method. 
Instead of directly sampling from the $\softmax$ distribution over $\gD\backslash\{d^+\}$ that inclining to the very top candidates, we follow the previously common practice to cap top-N (says N=200 in our experiments) candidates and then conduct a uniform sampling to ensure its diversity. 
As for sampling over the joint negative distribution in Eq.(\ref{eq:neg_distribution_final}), we combine the capped top-N candidates from multiple generators (retrievers) without de-duplication to better simulate the joint distribution.

\section{Experiments} \label{sec:exp}
\subsection{Datasets \& Metrics} 
In the experiment, we adopt the popular passage retrieval dataset, MS-Marco \citep{Nguyen16Tarek} for model training and evaluation. 
We utilize official queries on the MS-Marco dataset and report the model results on MS-Marco Dev \citep{Nguyen16Tarek} to report BM25-reranking, full-ranking and first-stage retrieval, and TREC Deep Learning 2019 \citep{Craswell20Overview} for our method. 
Following previous work, on the BM25-reranking and full-ranking task of the MS-Marco Dev dataset, we adopt MRR@10 to report the performance. On TREC Deep Learning 2019, we use NDCG@10 to report performance. 
In addition, following previous works \citep{Ren21RocketQAv2,Zhang21Adversarial}, we leverage our ranker to teach a retriever by knowledge distillation for first-stage retrieval, and evaluate the retrieval results using MRR@10 and R@50. The above MRR, NDCG, and R refer to Mean Reciprocal Rank, Normalized Cumulative Discount Gain, and Recall, respectively.

\begin{table}[t]\small
\centering
\begin{tabular}{lcccccc}\toprule
\textbf{Retriever}    & \textbf{\begin{tabular}[c]{@{}c@{}}BM25 \\ Reranking\end{tabular}} & \textbf{\begin{tabular}[c]{@{}c@{}}Den-BN \\ Reranking\end{tabular}} & \textbf{\begin{tabular}[c]{@{}c@{}}Den-HN \\ Reranking\end{tabular}} & \textbf{\begin{tabular}[c]{@{}c@{}}Lex-BN \\ Reranking\end{tabular}} & \textbf{\begin{tabular}[c]{@{}c@{}}Lex-HN \\ Reranking\end{tabular}} & \multicolumn{1}{l}{\textbf{Avg.}} \\\midrule
\hline
\multicolumn{7}{l}{\textit{Retriever acts reranker.}} \\
\hline
BM25               & 21.23                                       & 22.50                                     & 21.90                                     & 21.61                                     & 21.58                                     & 21.76                             \\
Den-BN (\textbf{abbr.} D1)                 & 35.51                                       & 36.15                                     & 36.14                                     & 36.28                                     & 36.24                                     & 36.06                             \\
Den-HN (\textbf{abbr.} D2)                 & 36.76                                       & 38.12                                     & 38.12                                     & 38.14                                     & 38.14                                     & 37.86                             \\
Lex-BN (\textbf{abbr.} L1)                 & 35.31                                       & 36.17                                     & 36.20                                     & 36.11                                     & 36.13                                     & 35.98                             \\
Lex-HN (\textbf{abbr.} L2)                 & 36.86                                       & 38.24                                     & 38.26                                     & 38.18                                     & 38.18                                     & 37.94                             \\\midrule
\hline
\multicolumn{7}{l}{\textit{Our ranker trained with retriever(s) as reranker.}} \\
\hline
\textsc{R$^2$anker}           &                                             &                                           &                                           &                                           &                                           &                                   \\
- BM25             & 39.82                                       & 41.38                                     & 41.44                                     & 41.39                                     & 41.41                                     & 41.09                             \\
- D1               & 40.50                                       & 42.81                                     & 42.82                                     & 42.78                                     & 42.82                                     & 42.34                             \\
- D2               & 40.47                                       & 42.62                                     & 42.67                                     & 42.60                                     & 42.62                                     & 42.20                             \\
- L1               & 39.81                                       & 41.56                                     & 41.56                                     & 41.92                                     & 41.77                                     & 41.32                             \\
- L2               & 40.78                                       & 41.51                                     & 41.60                                     & 42.92                                     & 42.88                                     & 41.94                             \\
- D1,D2            & 40.71                                       & 42.96                                     & 43.00                                     & 42.93                                     & 42.94                                     & 42.51                             \\
- L1,L2            & 40.44                                       & 42.19                                     & 42.03                                     & 42.93                                     & 42.81                                     & 42.08                             \\
- D1,L1            & 40.70                                       & 42.98                                     & 42.92                                     & 42.98                                     & 42.98                                     & 42.51                             \\
- D2,L2            & 40.82                                       & 42.88                                     & 42.92                                     & 42.89                                     & 42.92                                     & 42.49                             \\
- BM25,D2,L2       & \textbf{41.12}                              & \textbf{43.24}                            & \textbf{43.26}                            & \textbf{43.28}                            & \textbf{43.29}                            & \textbf{42.84}                    \\
- D1,L1,D2,L2      & 41.00                                       & 43.21                                     & 43.22                                     & 43.21                                     & 43.24                                     & 42.77                             \\
- BM25,D1,L1,D2,L2 & 40.78                                       & 42.99                                     & 42.95                                     & 42.97                                     & 42.99                                     & 42.54                             \\\bottomrule
\end{tabular}
\caption{\small Full-ranking results in terms of MRR@10 by different retriever-ranker combinations.}
\label{tab:all}
\end{table}

\subsection{Experimental Setups}
For the training of the ranker, we use the ERNIE-2.0-base model \citep{Sun19ERNIE} as the initialization of our ranker. To provide more diverse hard negatives for ranker’s robust training,  we sample them from multiple retrievers. In our experiments, we use three kinds of retrievers, including BM25 \citep{Yang17Anserini}, coCondenser \citep{Gao22Unsupervised} as dense-vector retrieval models and SPLADE \citep{Formal21SPLADEv2} as lexicon-weighting retrieval models. 
During ranker training, we sample 40 hard negatives for each query. The maximum training epoch, batch size and learning rate are set to 2, 12 and $1 \times 10^{-5}$. The maximum sequence length is set to 128 and the random seed is fixed to 64. For model optimizing, we use Adam optimizer \cite{Kingma14Adam} and a linear warmup. The warmup proportion is 0.1, and the weight decay is 0.1. All experiments are conducted on an A100 GPU.

To distill our trained ranker to a retriever for first-stage retrieval, we adopt the two-stage coCondenser retriever \citep{Gao22Unsupervised} and merely apply our ranker scores to the second stage. Specifically, instead of contrastive learning, we leverage training data by \citet{Ren21RocketQAv2} and discard contrastive learning loss as in coCondenser, but a simple KL divergence loss. Learning rate, batch size, and epoch number is set to $5\times10^{-5}$, $16 \times$ (1 positive and 10 negatives), and 4, respectively.

\subsection{BM25-Reranking and Full-Ranking}

First of all, we briefly introduce the involved baseline retrieval models in the following. 

\begin{itemize}
    \item {\bf BM25:} A default unsupervised retrieval method \citep{Yang17Anserini}.
    \item {\bf BERT$_{\text{base}}$ + ranking:} \citet{Qiao19Understanding} use a pretrained BERT model for ranking tasks.
    \item {\bf ColBERT:} The method employs deep LMs (in particular, BERT) to late interaction paradigm for efficient retrieval \citep{Khattab20ColBERT}.
    \item {\bf BERT:} A simple re-implementation of BERT for query-based passage re-ranking \citep{Nogueira19Passage}.
    \item {\bf SAN + BERT$_{\text{base}}$:} Using a stochastic answer network (SAN) and pre-trained language model for passage reranking \citep{Liu18Stochastic}.
    \item {\bf RocketQA:} The method proposes three training strategies, comprising cross-batch negatives, denoised hard negatives and data augmentation \citep{Qu21RocketQA}.
    \item {\bf Multi-stage:} Arranging monoBERT and duoBERT in a multi-stage ranking architecture to form an end-to-end search system \citep{Nogueira19Multi}.
    \item {\bf CAKD:} A cross-architecture training procedure with a margin focused loss (Margin-MSE) \citep{Sebastian20Improving}.
    \item {\bf RocketQAv2:} The method introduces the dynamic listwise distillation for unified training of both the retriever and the reranker \citep{Ren21RocketQAv2}.
\end{itemize}

\begin{table}[t]\small
\centering
\begin{tabular}{lccc}
\toprule
\textbf{Ranker}       & \textbf{Retriever} & \textbf{\#Cand\&Recall}  & \textbf{MRR@10} \\\midrule
\hline
\textit{BM25-Reranking} & & & \\
\hline
BM25 \citep{Yang17Anserini}  & BM25            & 1000 (85.7)               & 18.7            \\
BERT$_{\text{base}}$ \citep{Qiao19Understanding} & BM25            & 1000 (85.7)                & 33.7             \\
ColBERT \citep{Khattab20ColBERT}  & BM25            & 1000 (85.7)               & 34.9            \\
BERT$_{\text{large}}$ \citep{Nogueira19Passage} & BM25            & 1000  (85.7)               & 36.5            \\
SAN + BERT$_{\text{base}}$ \citep{Liu18Stochastic} & BM25            & 1000  (85.7)               & 37.0            \\
RocketQA \citep{Qu21RocketQA} & BM25            & 1000  (85.7)               & 37.0            \\
Multi-stage \citep{Nogueira19Multi} & BM25            & 1000  (85.7)               & 39.0            \\
CAKD \citep{Sebastian20Improving} & BM25            & 1000  (85.7)               & 39.0            \\
RocketQAv2 \citep{Ren21RocketQAv2} & BM25            & 1000  (85.7)               & 40.1            \\
\textsc{R$^2$anker} (Ours)  & BM25            & 1000  (85.7)               & \textbf{41.1}   \\\midrule 
\hline
\textit{Full-Ranking} & & & \\
\hline
RocketQA \citep{Qu21RocketQA}   & RocketQA \citep{Qu21RocketQA}  & 50 (85.5)                 & 40.9            \\
RocketQAv2 \citep{Ren21RocketQAv2} & RocketQA \citep{Qu21RocketQA}  & 50 (85.5)                  & 41.8            \\
RocketQAv2 \citep{Ren21RocketQAv2} & RocketQAv2  \citep{Ren21RocketQAv2} & 50 (86.2)                  & 41.9            \\
\textsc{R$^2$anker} (Ours) & coCondenser$*$ \citep{Gao22Unsupervised}  & 50 (83.5)                 & 42.7    \\
\textsc{R$^2$anker} (Ours) & coCondenser$\S$ \citep{Gao22Unsupervised}  & 50 (86.4)                 & \textbf{43.3}   \\
\textsc{R$^2$anker} (Ours) & SPLADEv2$*$ \citep{Formal21SPLADEv2}  & 50 (84.3)                 & 43.0    \\
\textsc{R$^2$anker} (Ours) & SPLADEv2$\S$ \citep{Formal21SPLADEv2}  & 50 (86.1)                 & \textbf{43.3}   \\\bottomrule
\end{tabular}
\caption{
BM25-reranking and full-ranking results on MS-Marco Dev. Note that all rankers in `\textit{full-ranking}' are built upon base pre-trained models. 
`\#Cand\&Recall' denotes the number of retriever-provided top candidates for reranking, as well as the retriever's top-N recall metric. 
$*$ denotes the retrieval model trained on BM25 negatives.
$\S$ denotes the retrieval model trained on hard negatives sampled by its $*$ retrieval model.}
\label{tab:results}
\label{tab:diff_retriever}
\end{table}

\begin{table}[t] \small
    \centering
    \begin{tabular}{lc}\toprule
    \textbf{Method} & \textbf{NDCG@10} \\\midrule
    RocketQAv2 \citep{Ren21RocketQAv2}     & 71.4             \\
    \textsc{R$^2$anker} (Ours)            & \textbf{73.0}    \\\bottomrule
    \end{tabular}
    \caption{BM25-reranking results on TREC Deep Learning 2019.}
    \label{tab:trec}
\end{table}

\paragraph{Comprehensive Full-Ranking on MS-Marco Dev.}
Table~\ref{tab:all} shows full-ranking results in terms of MRR@10 by different retriever-ranker combinations. From the table, we can make two observations: First, we can see that \textsc{R$^2$anker}-BM25,D2,L2 achieves the best performance over different retrieval results. Second, the results show that the performance of our ranker tends to increase by introducing hard negatives for more different retrievers. But introducing hard negatives from too many different retrievers, we found that the performance no longer improved and even had a tendency to decline. 
Please refer to the following discussion for more analyses regarding these observations.
In the subsequent discussion, for the convenience of analyzing our ranker, we employ \textsc{R$^2$anker} to denote \textsc{R$^2$anker}-BM25,D2,L2.

\paragraph{BM25-Reranking on MS-Marco Dev.}
The results of BM25-reranking on MS-Marco Dev are listed in Table \ref{tab:results}. As we can see, the results show that our method achieves a state-of-the-art performance, which demonstrates that our reranker is more robust. The reason is that our method introduces more diverse hard negatives from multiple retrievers, which can alleviate the noise learning inherent to the particular retriever by the ranker. In addition, we can observe that co-training methods (i.e., RocketQA, Multi-stage, CAKD, RocketQAv2) can significantly improve the performance of reranking. However, our method does not use co-training but introduces hard negatives from different retrievers and outperforms other co-training methods by about 1.0\% $\sim$ 4.1\% on MRR@10. It indicates that the reason for the co-training strategy improving the performance could be introducing more diverse hard negatives. The results also demonstrate the superiority of our method for training a more robust ranker without complex co-training of multiple models. It significantly reduces the training cost of the model.

\paragraph{BM25-Reranking on TREC Deep Learning 2019.}
To further verify the effectiveness of our method, we evaluate our approach and the strong baseline (RocketQAv2) on TREC Deep Learning 2019 dataset. As shown in Table \ref{tab:trec}, we can observe that our method significantly outperforms RocketQAv2 by 2.6 on NDCG@10. It further demonstrates the effectiveness of our approach. Meanwhile, it shows that introducing more diverse hard negatives can guide ranker training with better generalization.

\begin{table}[t]\small 
\centering
\setlength\tabcolsep{3pt}
\begin{tabular}{lccccc}\toprule
\textbf{Method}      & \textbf{PLM (Size)} & \textbf{Teacher (Size)}             & \textbf{Co-Training} & \textbf{MRR@10} & \textbf{R@50} \\\midrule
BM25 \citep{Yang17Anserini}                & -                   & -                                   &                      & 18.7            & 59.2          \\
doc2query \citep{Nogueira19Document}           & BERT$_{\text{base}}$                   & -                                   &                       & 21.5            & 64.4          \\
DeepCT \citep{Dai19Deeper}              & BERT$_{\text{base}}$                   & -                                   &                       & 24.3            & 69.0          \\
docTTTTTquery \citep{Cheriton19FromDT}       & T5$_{\text{base}}$                  & -                                   &                       & 27.7            & 75.6          \\
ANCE  \citep{Xiong21Approximate}                 & RoBERTa$_{\text{base}}$                   & -                                   &                       & 33.0            & -             \\
ColBERT \citep{Khattab20ColBERT}              & BERT$_{\text{base}}$                   & -                                   &                       & 36.0            & 82.9          \\
RocketQA \citep{Qu21RocketQA}             & ERNIE$_{\text{base}}$                  & ERNIE$_{\text{base}}$                                   & $\checkmark$                  & 37.0            & 85.5          \\
COIL  \citep{Gao21COIL}                  & BERT$_{\text{base}}$                   & -                                   &                       & 35.5            & -             \\
ME-BERT  \citep{Luan21Sparse}             & BERT$_{\text{large}}$                   & -                                   &                       & 33.8            & -             \\
PAIR  \citep{Ren21PAIR}               & ERNIE$_{\text{base}}$                   & -                                   &                       & 37.9            & 86.4          \\
DPR-PAQ  \citep{Oguz21Domain}   & BERT$_{\text{base}}$                   & -                                   &                       & 31.4            & -             \\
DPR-PAQ \citep{Oguz21Domain} & RoBERTa$_{\text{base}}$                   & -                                   &                       & 31.1            & -             \\
Condenser  \citep{Gao2021Condenser}          & BERT$_{\text{base}}$                   & -                                   &                   & 36.6            & -             \\
coCondenser \citep{Gao22Unsupervised}          & BERT$_{\text{base}}$                   & -                                  &                  & 38.2            & -             \\
RocketQAv2 \citep{Ren21RocketQAv2}           & ERNIE$_{\text{base}}$                   & ERNIE$_{\text{base}}$                                   & $\checkmark$                  & 38.8            & 86.2          \\ \hline
\textbf{Ours}                 &  coCondenser$_{\text{b}}$                  & \textsc{R$^2$anker}$_{\text{base}}$ &     & \textbf{40.0}   & \textbf{87.6}    \\
\midrule \hline
AR2  \citep{Zhang21Adversarial}                  & coCondenser$_{\text{b}}$                   & ERNIE$_{\text{large}}$                                   & $\checkmark$                  & 39.5            & \textbf{87.8} \\
ERNIE-Search \citep{Lu22ERNIE}                  & ERNIE$_{\text{base}}$                   & ERNIE$_{\text{large}}$                                   & $\checkmark$                  & \textbf{40.1}            & 87.7 \\
\bottomrule
\end{tabular}
\caption{\small First stage retrieval performance on MS-Marco Dev. A non-empty `Teacher' denotes that the retriever is trained with a ranker, and `Co-Training' denotes the ranker is also updated during training, otherwise frozen.}
\label{tab:retrieval}
\end{table}

\paragraph{Full-Ranking with Different Retrievers.}
To more comprehensively verify our claim and compare our method and other baselines, we compare our method with other strong baselines under the different retriever. The results are shown in Table \ref{tab:diff_retriever}. From the results, we can see that our method outperforms other models by about 1.4\% $\sim$ 2.4\% on MRR@10. It is worth noting that since our method only trains a reranker, we need to use other retrievers to provide a preliminary retrieval result first. Moreover, we can observe that our ranker still achieves better performance, even though our method is based on a weaker retriever whose performance underperforms RocketQAv2 (i.e., coCondenser$*$ \citep{Gao22Unsupervised} is 83.5 and RocketQAv2 is 86.2  on R@50). Therefore, our model can achieve better reranking performance even on a sub-optimal retrieval result. 

\subsection{Knowledge Distillation for Retriever}
Under the `retrieval \& rerank' pipeline, the ranker can often rerank the results from the retriever to get better full-ranking results. 
Some recent works use ranker to guide retriever training, specifically by distilling the reranker's scores to the retriever, which has been shown to be very effective  \citep{Ren21RocketQAv2,Zhang21Adversarial}. To verify that our ranker learns more robust knowledge, we distill our ranker to a bi-encoder retriever. 
We compare the trained retriever with recently advanced competitors, and the results are shown in Table \ref{tab:retrieval}. 
It is observed that, with the comparable ranker (teacher) size, our model achieves state-of-the-art performance in terms of both MRR@10 and R@50. Meantime, our distilled retriever's performance even keeps competitive with the retrievers with larger (i.e., $3\times$) teachers. 
Moreover, it is noteworthy that our distillation paradigm is very easy, i.e., no need for co-training or adv-learning, and can be plug-in into any retriever training with reduced computation overheads.

\subsection{Distribution Analysis} 
\begin{figure}[t]
    \centering
    \includegraphics[width=0.65\linewidth]{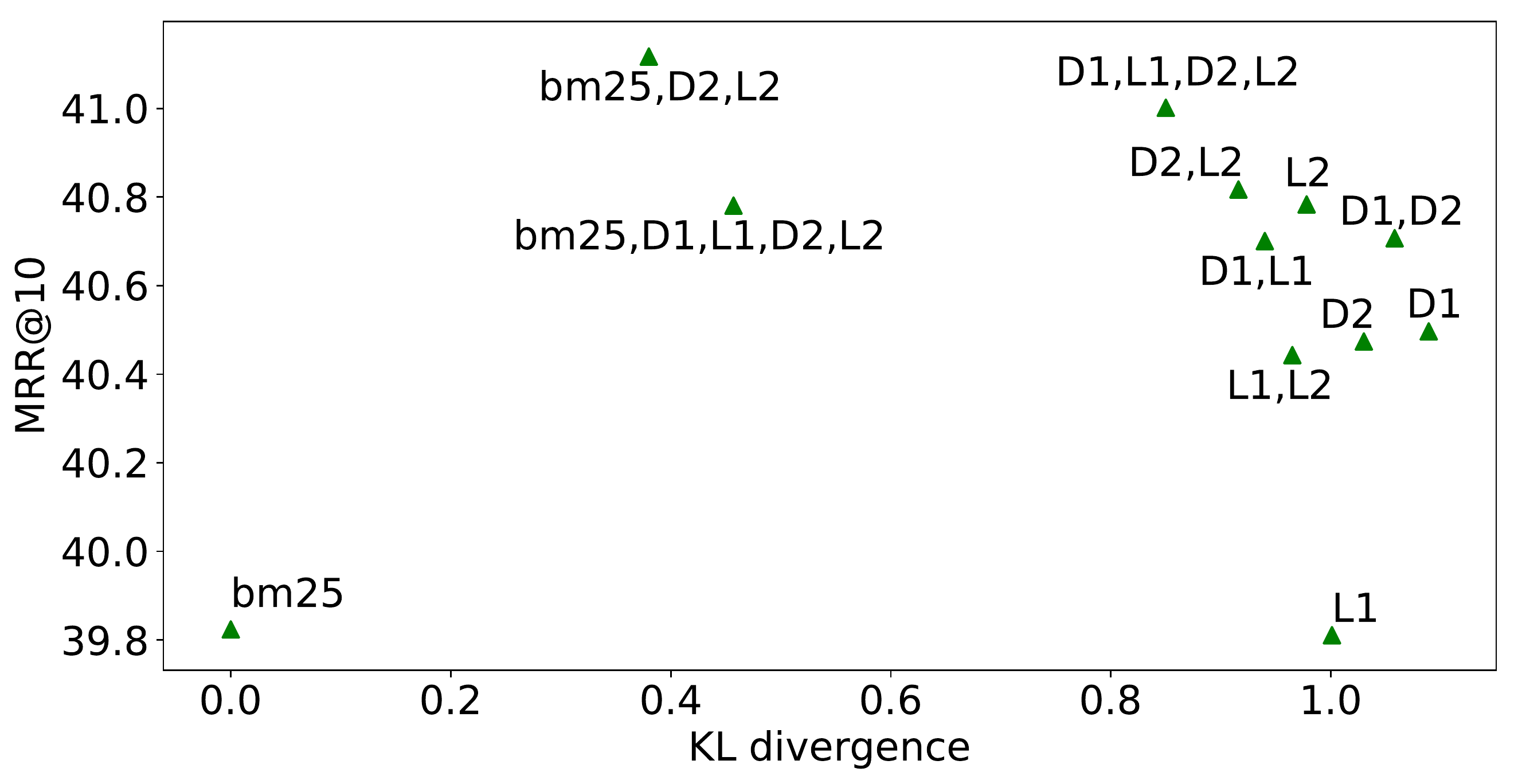}
    \caption{\small 
    BM25-reranking performance by various rankers that were trained on negatives sampled from retrievers' (joint) distributions. 
    D1, D2, L1, and L2 is abbreviations for Den-BN, Den-HN, Lex-BN, and Lex-HN retriever, respectively.
    `KL divergence' denotes the difference between the retrievers' (joint) distribution and BM25 retriever's, i.e., $KL(P(\cdot|q; \Theta^{\text{(be)}})|\bmsample(\cdot|q; \gD))$, which is used to measure negatives' distribution.
    For example, the point `bm25,D2,L2' denotes that i) the KL between its joint retriever's distribution and BM25 retriever's distribution is round 0.4, and ii) a ranker trained on that joint negative distribution can achieve 41.1 MRR@10 on BM25 reranking. 
    }
    \label{fig:analysis}
\end{figure}

\paragraph{Distribution of Hard Negative.}
As shown in Figure~\ref{fig:analysis}, we show BM25-reranking performance for the rankers that were trained on negatives sampled from retrievers' (joint) distributions, $P(\cdot|q; \Theta^{\text{(be)}})$. 
Here, we assume that the negatives sampled by the BM25 retriever are too moderate to train a robust ranker. 
Thereby, we propose to measure the diversity of negatives from $P(\cdot|q; \Theta^{\text{(be)}})$ by a Kullback–Leibler (KL) divergence between $P(\cdot|q; \Theta^{\text{(be)}})$ and $\bmsample(\cdot|q; \gD))$. 
It is expected that, the more dissimilar distribution to BM25, the more diverse negative samples, making the performance better. 
However, as shown in the figure, the most effective joint retriever involves BM25.
A potential reason is that, the almost non-parametric BM25 is orthogonal to the other learnable retrievers (both dense-vector and lexicon-weighting ones), making the negative diverse enough for a great ranking quality.

\begin{figure}[t]
    \centering
    \includegraphics[width=0.65\linewidth]{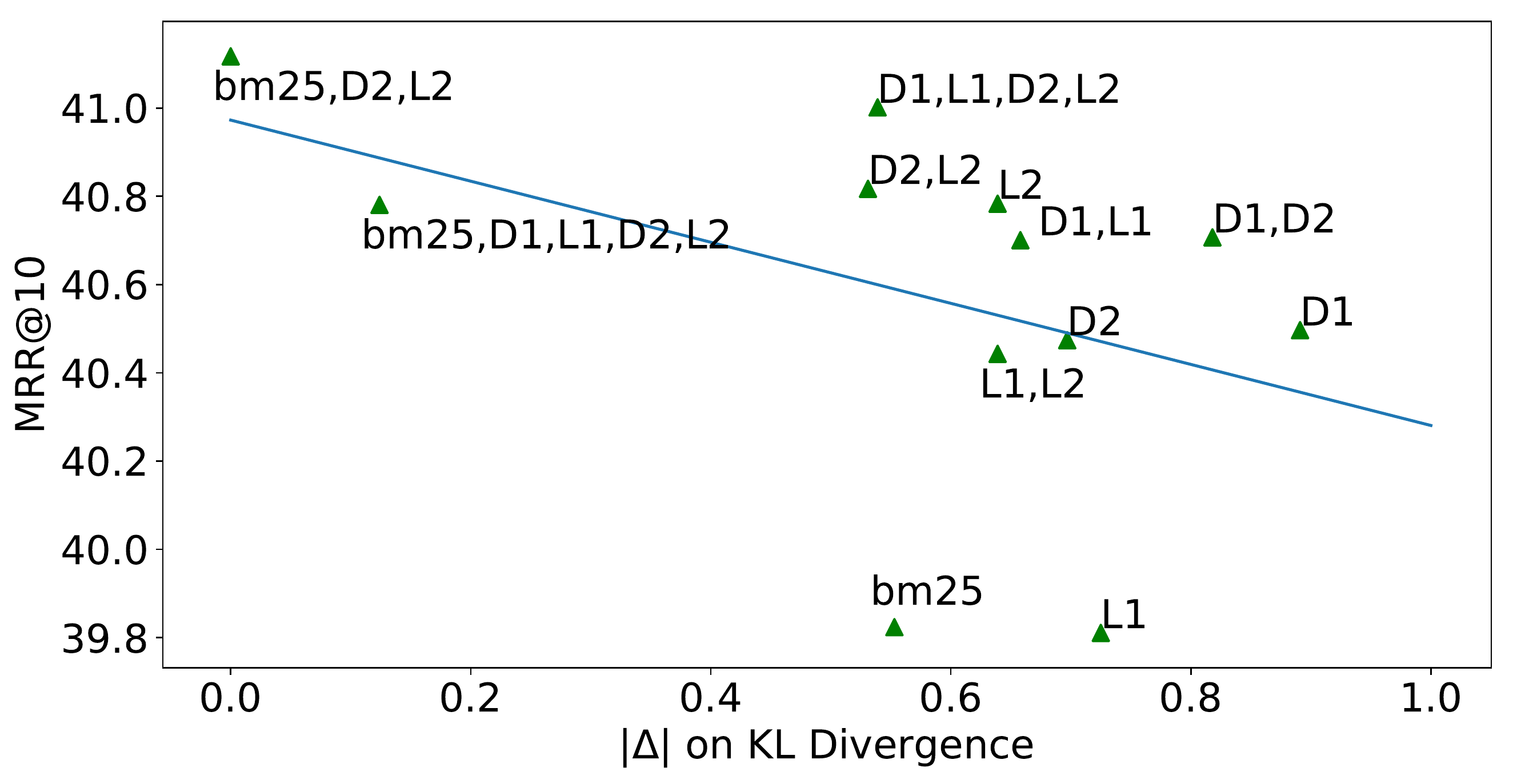}
    \caption{
    BM25-reranking performance by various rankers \textit{vs.} relevance score distribution changes from the (joint) retriever to the trained ranker. 
    In formal, $\Delta = KL(P(\cdot|q; \Theta^{\text{(be)}})|\bmsample(\cdot|q; \gD)) - KL(P(\cdot|q; \theta^{\text{(ce)}})|\bmsample(\cdot|q; \gD))$, where $\theta^{\text{(ce)}}$ is trained with a specific $\Theta^{\text{(be)}}$.
    The smaller $|\Delta|$, the negative sampling distribution closer to the ideal negative distribution of the ranker, as learning on the retriever-sampled negatives will not shift the  distribution. 
    }
    \label{fig:compare}
\end{figure}

\paragraph{Distribution Shift. }

We want to investigate whether the distribution shifted from a (joint) retriever to a correspondingly trained ranker will correlate to the ranker's performance. 
We assume that, a smaller distribution shift represents that the joint retriever is closer to the ideal negative distribution of a ranker, which is an echo of our motivations and claims. 
As shown in Figure~\ref{fig:compare}, we illustrate the reranking performance w.r.t. $|KL(P(\cdot|q; \Theta^{\text{(be)}})|\bmsample(\cdot|q; \gD)) - KL(P(\cdot|q; \theta^{\text{(ce)}})|\bmsample(\cdot|q; \gD))|$, which verifies our assumption.

\section{Related Work} \label{sec:rel_work}
\subsection{Information Retrieval}
In information retrieval, `retrieval \& rerank' has become a standard and default pipeline \citep{Guo22Semantic}. The reason is twofold: First, it is not feasible to train on the entire collection; Second, human annotation is costly. For the retrieval stage, given a textual query, a retriever can return a relevant score between it and each textual item (e.g., paragraphs and documents) from large-scale collections. The retriever is generally implemented by a bi-encoder (also known as dual encoder and siamese encoder) because they can independently derive representations of texts and compute correlations between texts efficiently. Then, the top-k retrieval results derived from the retriever are passed to the reranker to obtain more accurate ranking results. Therefore, the reranker is a crucial part of the pipeline and directly affects the final performance of information retrieval. Meanwhile, it has been attracting the attention of more researchers \citep{Qu21RocketQA,Ren21RocketQAv2,Zhang21Adversarial}. In this work, we focus on reranker learning due to its importance.

In addition, rerankers are currently widely used as a teacher in retriever training. The scores derived by the reranker are demonstrated that they can guide the retriever learning through knowledge distillation \citep{Ren21RocketQAv2,Zhang21Adversarial}. RocketQAv2 \citep{Ren21RocketQAv2} introduces a joint training approach for dense passage retrieval and passage reranking, which uses dynamic listwise distillation to dynamically update both the parameters of the reranker and the retriever. AR2 \citep{Zhang21Adversarial} presents an adversarial framework for dense retrieval, where the retriever is regarded as the generator, and the ranker is viewed as the discriminator. Therefore, the reranker can not only be further used to rank the results from the retriever but also improve the performance of the retriever through knowledge distillation.

There is a discrepancy between the training and inference of the retriever. The reason is that candidate passages annotated for one question are from the top-K passages retrieved by a specific retrieval approach (e.g., BM25 \citep{Yang17Anserini}). The rest of the collection may still have positive passages. During the retriever training, the model is learned to estimate the probabilities of positive passages in a small candidate set for each question, which leads to difficulty in effectively training a solid retriever. The training of the reranker is based on the hard negatives retrieved by the retriever but is full of false negatives in them. Therefore, an imperfect retriever and noisy data also lead to suboptimal performance of the reranker.

\subsection{Learning with Noise Data}
To alleviate the discrepancy between the training and inference of the retriever, some works \citep{Qu21RocketQA,Zhang21Adversarial} propose reselecting some hard negatives from the retrieval results. Although this method dramatically improves retrieval performance, it still suffers from false negatives. To remove false negatives from the top-ranked results retrieved by a retriever, \citet{Qu21RocketQA} introduce denoised hard negatives, using a cross-encoder based reranker to remove top-retrieved passages that are likely to be false negatives. \citet{Zhang21Adversarial} construct a unified minimax game for training the retriever and ranker interactively to reduce the effect of false negatives. Although these methods can mitigate the effect of false negatives for retriever, false negatives retrieved by the retriever still affect the training of the reranker.

In addition, many other works in learning with noise data \citep{Han18Coteaching,Wei21Open} show remarkable results. Some works by weighting or filtering the training samples, such as small-loss selection \citep{Han18Coteaching,Wei20Combating}, (dis)agreement between two models \citep{Malach17Decoupling,Wei20Combating}, or GMM distribution \citep{Arazo19Unsupervised,Li20DivideMix}. Moreover, there are some works propose to improve generalization under label noise settings by regularization techniques, such as virtual adversarial training \citep{Miyato19Virtual}, gradient clipping \citep{Menon20Can}, label smoothing \citep{Lukasik20Does,Szegedy16Rethinking} and temporal ensembling \citep{Laine17Temporal}. Recently, some works attempt to alleviate the effect of data noise by introducing open-set noise that obeys different distributions \citep{Wei21Open}. Intuitively, based on the ``\textit{insufficient capacity}'' assumption \citep{Arpit17Closer}, increasing the number of open-set auxiliary samples slows down the fitting of inherent noises. In the `retrieval \& rerank' pipeline, the training of the reranker relies on the hard negatives obtained by the retriever. Still, hard negatives are inevitably full of false negatives, which spoil the performance of the reranker. Therefore, in this work, we focus on how to train a robust reranker on noise data.

\subsection{Learning with Multi-generator}
Recently, learning with multi-generators has been proven to improve models significantly and has attracted much attention \citep{Zhou21Triple,Zhang19EnsembleGAN,Hoang18MGAN}. \citet{Zhou21Triple} propose a triple adversarial framework for unsupervised caption generation, which comprise an image generator, a sentence generator and a discriminator. The framework can achieve aligning representations across modalities through an adversarial game between generators and the discriminator. \citet{Zhang19EnsembleGAN} propose an adversarial learning framework, ensembleGAN, consisting of a language-model-like generator, a ranker generator, and a discriminative ranker. In the framework, the generator and discriminator improve each other through the generation and retrieval-based methods. Although these co-training methods are effective, the model suffers from high training costs. In this work, instead of co-training to improve model performance, we train a more robust model by sampling diverse training samples from different models.

\section{Conclusion}
In this work, we present a simple yet effective learning strategy to reach a robust ranker, \textsc{R$^2$anker}. The robustness is achieved by two aspects -- open-set label noises making the ranker against inherent label noises and a multi-retriever joint negative sampling distribution close to the negative distribution of a ranker. 
Upon experiments on passage retriever, our proposed method achieve state-of-the-art performance on both BM25-reranking and full-ranking settings.
In addition, our trained model can serve as a teacher to learn a strong retriever, which can achieve very competitive results on first-stage retrieval. 
Lastly, extensive analytical experiments verify the correctness of our assumptions and claims.

\bibliography{reference}
\bibliographystyle{iclr2022_conference}

\end{document}